\documentstyle[prl,aps,epsf,epsfig,floats,twocolumn]{revtex}
\begin{document}
\draft
\twocolumn[\hsize\textwidth\columnwidth\hsize\csname
@twocolumnfalse\endcsname

\title{
\hfill{\small { FZJ-IKP(TH)-2000-29}}\\[0.2cm]
The S--wave $\Lambda\pi$ phase shift is not large}
%\vspace{1cm} 

\author{
Ulf-G. Mei{\ss}ner, Jos\'e Antonio Oller}

%\vspace{0.5cm}
\address{Forschungszentrum J\"ulich, Institut f\"ur Kernphysik (Th), D-52425 
 J\"ulich, Germany }
\maketitle
%\thispagestyle{empty}

%\vspace{2cm}

\begin{abstract}
We study the strong interaction S--wave $\Lambda\pi$ phase shift in the region
of the $\Xi$ mass in the framework of a relativistic chiral unitary
approach based on coupled channels. All parameters have been
previously determined in a fit to strangeness $S= -1$ S--wave
kaon--nucleon data. We find $0^\circ \le \delta_0 \le 1.1^\circ$ in agreement
with previous chiral perturbation theory calculations (or extensions
thereof). We also discuss why a recent coupled channel K-matrix calculation gives a
result for $\delta_0$ that is negative and much bigger in magnitude.
We argue why that value should not be trusted.
\end{abstract}

] 

\vspace{1cm}

\noindent
{\bf 1.}~Direct CP violation can be measured in the decay $\Xi \to \Lambda \pi
\to p\pi\pi$ (for a recent experiment, see \cite{luk}).
To extract the CP violating phase, one has to know the
strong $\Lambda \pi$ S-- and P--wave phase shifts at the mass of the
cascade, denoted $\delta_0$ and $\delta_1$, respectively. 
While earlier calculations~\cite{old1,old2} were inconclusive on the value of $\delta_0$,
a leading order heavy baryon chiral perturbation theory (HBCHPT) analysis led to 
a vanishing S--wave phase shift~\cite{LWS} and corrections including
excited $\Sigma$ intermediate states were shown to give a bound of
$\delta_0 \sim 0.5^\circ$~\cite{LWS,DP}. Relativistic tree level
calculations have also been performed, leading to a somewhat larger
band of values for $\delta_0$, but still $|\delta_0| \le 2^\circ$~\cite{DDP,K}.
A more recent calculation using also
dimension two operators~\cite{TTV} with the corresponding low--energy constants
fixed from kaon--nucleon scattering~\cite{KSW} gave the range
$-3.0^\circ \le \delta_0 \le +0.4^\circ$.\footnote{Note that
  the parameters obtained in \cite{KSW} need to be taken with some
  care since the important $\eta$ channels were not considered, as
  stressed in \cite{OR}.} 
In that paper, the effect of channel coupling was also investigated, based on the
observation that in SU(3), the $\Lambda \pi$ state is coupled to the
$\Sigma \pi$, $N\bar{K}$, $\Sigma\eta$ and $\Xi K$ states with
strangeness $S=-1$ and isospin $I=1$. A K-matrix approach was used
to calculate the channel coupling effects and a surprisingly
large $\delta_0 \simeq -7^\circ$ was found. The authors of
ref.\cite{TTV} have been
careful to point out that more refined coupled channel calculations
based on chiral perturbation theory
(CHPT) are necessary to further clarify this surprising result. 
We have recently presented  a 
novel relativistic chiral unitary approach based on coupled channels~\cite{OM}. 
Dispersion relations are used to perform the necessary resummation of 
the lowest order relativistic chiral Lagrangian. Within this
framework, the S--wave kaon--nucleon interactions for strangeness $S=-1$
were studied and a good description of the 
data  in the $K^- p$, $\pi \Sigma$ and $\pi \Lambda$ channels
(cross sections, threshold ratios, mass distribution in the
region of the $\Lambda(1405)$) was obtained.  This method can be systematically extended 
to higher orders, emphasizing its applicability to any 
scenario of strong self--interactions where the perturbative 
series diverges even at low energies. It is straightforward to
project out the $\Lambda \pi \to \Lambda \pi$ amplitude from our
coupled channel solutions and extract in a parameter--free manner
the corresponding S--wave phase shift. This is done here. To close the
introduction, we remark that our approach can also be used to
calculate the P--waves. Since there is no discrepancy in the
corresponding predictions for $\delta_1$, we focus here entirely
on the S--wave.

\medskip\noindent
{\bf 2.}~We briefly summarize our calculational scheme, for details see \cite{OM}.
It is  based on the fact that unitarity, above the pertinent 
thresholds, implies that the inverse of a partial wave amplitude satisfies
\begin{equation}
\label{uni}
\hbox{Im}~T^{-1}(W)_{ij}=-\rho(W)_i \delta_{ij}~,
\end{equation}
where $\rho_i \equiv q_i/(8\pi W)$, $W = \sqrt{s}$ the centre-of-mass (cm)
energy, $q_i$ is the modulus of the cm 
three--momentum and the subscripts $i$ and $j$ refer to the physical channels. 
The $\Lambda \pi$ states couple strongly to several 
channels. To be consistent with lowest order CHPT, where all the baryons belonging 
to the same SU(3) multiplet are degenerate, one should consider the whole set of states: 
$K^-p~(1)$, $\bar{K}^0 n~(2)$, $\pi^0 \Sigma^0~(3)$, $\pi^+ \Sigma^-~(4)$, 
$\pi^- \Sigma^+~(5)$, $\pi^0 \Lambda~(6)$, $\eta \Lambda~(7)$, $\eta \Sigma^0~(8)$, 
$K^+\Xi~(9)$, $K^0 \Xi^0~(10)$, where 
between brackets the channel number, to be used in a matrix notation, is given for each 
state. The unitarity relation in eq.(\ref{uni}) gives rise to a cut in the
$T$--matrix of partial wave amplitudes which is usually called the unitarity or right--hand 
cut. Hence we can write down a dispersion relation for $T^{-1}(W)$, in a fairly symbolic 
language: 
\begin{eqnarray}
\label{dis}
T^{-1}(W)_{ij} &=&-\delta_{ij}\left\{\widetilde{a}_i(s_0) 
+ \frac{s-s_0}{\pi}\right. \nonumber \\ &\times& \left. \int_{s_{i}}^\infty ds' 
\frac{\rho(s')_i}{(s'-s)(s'-s_0)}\right\}+{\mathcal{T}}^{-1}(W)_{ij} ~,
\end{eqnarray}
where $s_i$ is the value of the $s$ variable at the threshold of channel $i$ and 
${\mathcal{T}}^{-1}(W)_{ij}$ indicates other contributions coming from local and 
pole terms as well as crossed channel dynamics but {\it without} 
right--hand cut. These extra terms will be  taken directly from CHPT 
after requiring the {\em matching} of our general result to the CHPT expressions. 
Notice also that the negative of the quantity in the curly brackets, denoted $g(s)_i$
from here on, is the familiar scalar loop integral
\begin{eqnarray}
\label{g2}
g(s)_i&=&{i}\int \frac{d^4 q}{(2\pi)^4}\frac{1}{(q^2-M_i^2+i \epsilon)
((P-q)^2-m_i^2+i\epsilon)}\nonumber \\
&=&\frac{1}{16 \pi^2}\left\{ a_i(\mu)+\log\frac{m_i^2}{\mu^2}+
\frac{M_i^2-m_i^2+s}{2 s}\log\frac{M_i^2}{m_i^2} \right.\nonumber\\
&+&\left.\frac{q_i}{\sqrt{s}}\log\frac{m_i^2+M_i^2-s-2 
\sqrt{s}q_i}{m_i^2+M_i^2-s+2\sqrt{s}q_i} \right\} ,
\end{eqnarray}
where $M_i$ and $m_i$ are, respectively, the 
meson and baryon masses in the state $i$. Notice that in order to calculate $g(s)_i$, 
we are using the physical masses both for mesons and baryons since the unitarity 
result in eq.(\ref{uni}) is exact. In the usual chiral power counting,  
$g(s)_i$ is ${\mathcal{O}}(p)$ because the baryon propagator scales as 
${\cal O}(p^{-1})$. Let us note that the important 
point here is to proceed systematically guaranteeing that ${\cal T}$ is
free of the right--hand cut  and matching simultaneously with the CHPT
expressions.  We can further simplify the notation by employing a matrix formalism. We  
introduce the 
matrices $g(s)={\rm diag}~(g(s)_i)$, $T$ and ${\mathcal{T}}$, the latter defined in 
terms 
of the matrix elements $T_{ij}$ and ${\mathcal{T}}_{ij}$. In this way, 
from eq.(\ref{dis}), the $T$-matrix can be written as:
\begin{equation}
\label{t}
T(W)=\left[I+{\mathcal{T}}(W)\cdot g(s) \right]^{-1}\cdot {\mathcal{T}}(W)~.
\end{equation}
In this short note, we are considering the lowest order (tree level) CHPT amplitudes as 
input. Hence,  expanding the previous equation,  our final expression for the 
$T$-matrix, taking as input the lowest order CHPT results, has the form
\begin{equation}
\label{fin}
T(W)=\left[ I + T_1(W) \cdot g(s) \right]^{-1} \cdot T_1(W) ~.
\end{equation}
For more details on this formalism, we refer to refs.\cite{OM,MO}. We only want to
remark that this approach is not just a unitarization scheme, like e.g. the
K--matrix approach. The latter is, however, included as one particular approximation
as discussed below.

\medskip\noindent
{\bf 3.}~Using the lowest order relativistic (tree level) CHPT amplitudes for $\phi_i B_a
\to \phi_j B_b$ as input, where $\phi_i \, (B_a)$ denotes a member
 of the Goldstone boson (ground state baryon) octet, one obtains
a very good description of the scattering data for $K^- p \to
K^-p, K^0 n, \pi^+\Sigma^-,  \pi^-\Sigma^+, \Lambda \pi^0, \Sigma^0
\pi^0$ (for kaon lab momenta below 250~MeV), 
the so--called threshold ratios $\gamma$, $R_c$ and $R_n$,
the $K^- p$ scattering length and the $ \pi^+\Sigma^-$ event
distribution in the region of the $\Lambda (1405)$ in terms of
three parameters (using fixed axial couplings, $D=0.80$ and $F=0.46$~\cite{rat}). 
These are the baryon octet mass in the chiral limit,
$m_0$, the chiral limit value of the three--flavor meson decay
constant\footnote{We remark that there are some indications that the
order parameter of chiral symmetry breaking, $F_0$, decreases sizeably
when going from the two to the three--flavor sector, see
e.g.~\cite{orsay}.}, 
$F_0$, and the subtraction constant $a(\mu)$, cf. eq.(\ref{g2}). 
Note that it was shown in \cite{OM} that it suffices to take only
one subtraction constant for {\em all} channels, thus the subscript
``$i\,$'' appearing in eq.(\ref{g2}) for these constants will be dropped out. 
In ref.\cite{OM}, we considered
two sets of parameters, set~I describing the best fit and set~II using
the so--called natural values (as discussed in that paper). The
pertinent numbers are for
set~I:~$m_0 =1.286\,$GeV, $F_0 = 74.1\,$MeV, $a(\mu) = -2.23\,$, and for
set~II:~$m_0 =1.151\,$GeV, $F_0 = 86.4\,$MeV, $a(\mu) = -2\,$ at the
scale $\mu =630\,$MeV. Of course, physical observables are
scale--independent. It is now straightforward to extract the
$\Lambda\pi$ phase shift as shown in fig.~1 by the solid line (set~I)
and the dashed line (set~II). The corresponding phases at the mass of
the $\Xi^0$ and the $\Xi^-$ are:
\begin{eqnarray}
{\rm set~I}: && \delta_0 (m_{\Xi^0}) = 0.10^\circ~, 
\delta_0 (m_{\Xi^-}) = 0.16^\circ~, \nonumber \\
{\rm set~II}: && \delta_0 (m_{\Xi^0}) = 0.92^\circ~, 
\delta_0 (m_{\Xi^-}) = 1.11^\circ~,
\end{eqnarray}
consistent with earlier CHPT findings~\cite{LWS,DP,DDP,K,TTV}. We should
stress that set~I gives the better fit in the $\bar{K}N$ sector and
should be preferred. 
\begin{figure}[htb]
%\vspace{-10cm}
\centerline{\epsfig{file=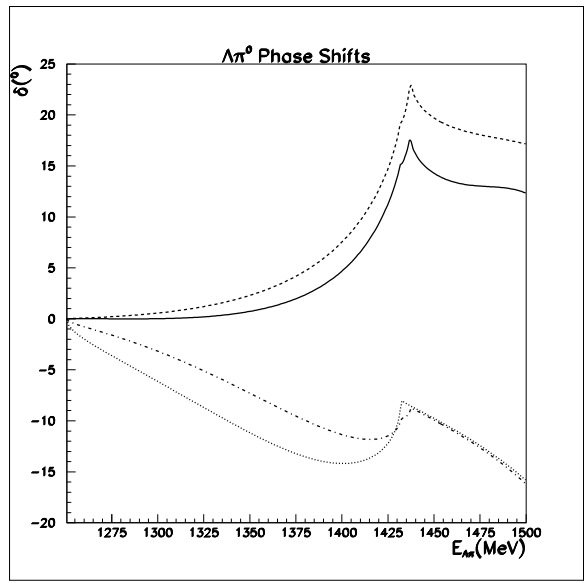,width=3.0in}}

\vspace{0.3cm}

\caption[pilf]{\protect \small
The $\Lambda \pi$ phase shift in degrees versus the cm energy,
$W= E_{\Lambda\pi}$. The various lines are explained in the text.
}
\end{figure}

\noindent It is important to understand the large result obtained
in the K--matrix formalism~\cite{TTV}. The K--matrix approach
is one particular approximation to our scheme in that ones sets
\begin{equation}
\label{ko}
g(s)_i = -\frac{i \,q_i}{8\pi W}\equiv -i\,\rho(s)_i~.
\end{equation}
Notice that $-\rho(s)_i$, above the threshold 
of channel $i$, is the imaginary part of $g(s)_i$, cf. eq.(\ref{dis}). In order to see 
the importance 
of keeping the whole $g(s)_i$ function, compare the 
dashed and dotted-dashed lines in fig.1. The latter is obtained for set~II by making 
use of eq.(\ref{fin}) but using the approximation given in eq.(\ref{ko}) to 
the $g(s)_i$ function. The 
differences are huge and for the second case the results are similar to the findings of 
ref.\cite{TTV}. In fact, we can 
reproduce the results for their K--matrix calculation by means of eq.(\ref{fin}) 
by considering only the dominant 
non--relativistic seagull (Weinberg-Tomozawa) term to the tree level meson--baryon 
scattering and the K--matrix representation of the $g(s)_i$ function. This  is given 
by the dotted line in fig. 1. All these large differences nicely show that 
it is not sufficient to account only for the imaginary part of the scalar 
loop functions via unitarity but that a proper treatment of the real part by an
appropriate dispersion relation is of equal importance. 
Consequently, the large and negative value for 
$\delta_0\simeq -7^\circ$ of ref.\cite{TTV} can be ruled out and is just a result of 
the simple 
representation of the function $g(s)_i$ used in that reference. This is, by far,  not 
sufficiently accurate for this case and the full relativistic expression for $g(s)_i$, 
cf. eq.(\ref{g2}), has to be used. Furthermore, the
phases are  sensitive to $F_0$ and $m_0$. We conclude from our approach  that indeed 
$\delta_0$ is narrowly bounded,
\begin{equation}\label{band}
0^\circ \le \delta_0 \le 1.1^\circ~,
\end{equation}
and that the large value found in the K--matrix approach should not be
used.

\medskip\noindent
{\bf 4.}~In summary, we have used a relativistic  chiral unitary approach based
on coupled channels to investigate the strong S--wave $\Lambda \pi$
phase shift in the region of the $\Xi$. All parameters have been
previously determined from a good description of the kaon--nucleon
data~\cite{OM} and thus we arrive at a small band of values for $\delta_0$, cf.
eq.(\ref{band}). This number is consistent with earlier findings in
CHPT (or extensions thereof)~\cite{LWS,DP,DDP,K,TTV}. We have also shown why
the K--matrix approach of ref.\cite{TTV} leads to a large value of
$\delta_0$ and why this number should not be trusted. The strong
$\Lambda\pi$ S--wave phase in the region of the cascade mass is indeed small.

\pagebreak

\section*{Acknowledgments}
The work of J.A.O. was supported in part by funds from
DGICYT under contract PB96-0753 and from the EU TMR network Eurodaphne, contract
no. ERBFMRX-CT98-0169.

%\vspace{-0.2cm}
%%%%%%%%%%%%%%%%%%%%%%%%%%%%%%% refs %%%%%%%%%%%%%%%%%%%%%%%%%%%%%%%%%%%%%

\end{document}